\documentclass[aps,prl,twocolumn,showpacs,superscriptaddress]{revtex4-1}
\usepackage{amsmath}	
\usepackage{color}
\usepackage{gensymb}
\usepackage{graphicx}

\usepackage{hyperref}
\usepackage{upgreek}
\usepackage{ulem}
\usepackage{MnSymbol}
\usepackage{bm}

\begin{document}

\title{Full Stark Control of Polariton States on a Spin-Orbit Hypersphere}

\author{Feng Li}	
\author{E. Cancellieri}
\email[]{Corresponding author: e.cancellieri@sheffield.ac.uk}
\author{G. Buonaiuto}
\author{M. S. Skolnick}
\author{D. N. Krizhanovskii}
\author{D. M. Whittaker}
\affiliation{Department of Physics and Astronomy, University of Sheffield, Sheffield S3 7RH, UK}
\date{\today}

\begin{abstract}
The orbital angular momentum and the polarisation of light are physical quantities widely investigated for classical and quantum information processing. In this work we propose to take advantage of strong light-matter coupling, circular-symmetric confinement, and transverse-electric transverse-magnetic splitting to exploit states where these two degrees of freedom are combined. To this end we develop a model based on a spin-orbit Poincar\'{e} hypersphere. Then we consider the example of semiconductor polariton systems and demonstrate full ultrafast Stark control of spin-orbit states. Moreover, by controlling states on three different spin-orbit spheres and switching from one sphere to another we demonstrate the control of different logic bits within one single physical system.
\end{abstract}

\maketitle

The polarisation of photons and the spin of photon-emitters, such as atoms, quantum dots and vacancy-defect centres, are among the most exploited physical properties for the implementation of classical as well as quantum information processing \cite{Lodahl2015,Carter2013,Vora2015,Holleczek11,Yao2011}. In recent years, considerable efforts have been devoted to the use of structured light beams with orbital angular momentum (OAM) to maximise information processing capabilities. Significant quantum effects such as entanglement of multi-photon states with high values of OAM and OAM Hong-Ou-Mandel interference have been demonstrated \cite{Yao2011,Mair2001,Fickler2012,Malik2015,Wang2015,Karimi2014}. The next natural step is to use higher-dimensional Hilbert spaces, like for example Spin-Orbit (SO) coupled states \cite{Prati1997,Cardano2012,Gong2014,Fickler20141,Milione2012}, which might allow simplify quantum logic \cite{multipleqbits}.

The strong coupling of photons with photon-emitters leads to the formation of polaritons, new half-light half-matter dressed states. A particular advantage offered by these hybrid quasiparticles is that they allow not only ultrafast manipulation through their light component \cite{LightSciAppl} but also through their matter component, opening the way to a more extended and flexible control. This can be achieved taking advantage of the AC Stark effect that allows controlling the excitation energy of photon emitters \cite{PhysRevB.65.155206,0953-8984-2-27-005} without modifying the population. This effect, recently been demonstrated for semiconductor microcavities \cite{PhysRevLett.109.033605,PhysRevLett.112.053601}, but can in principle also be applied to other systems such as: semiconductor or colloidal quantum dots and defect centres.

In this {\it Letter} we develop a theoretical model based on a SO hypersphere \cite{Agarwal:99} and use red-detuned laser pulses to manipulate angular momentum and polarisation of polariton states. This model has the unique advantage of combining multiple logical bits in one single physical system, and of allowing them to be independently manipulated. For the sake of clarity we will limit our analysis to the case of OAM $l=\pm 1$, but the theory can be generalised to higher values of $|l|$.

Our underlying general theory is valid for emitters in strong coupling with light in the presence of circular-symmetric confinement and transverse-electric transverse-magnetic (TE-TM) splitting. Parameters for state of the art semiconductor microcavity systems are used as an example to demonstrate the feasibility of our theoretical approach. Semiconductor polaritons are particularly interesting since they are reaching the maturity for quantum information processing \cite{PhysRevA.88.042310,PhysRevLett.112.196403} and because already allowed the observation of quantised vortices \cite{Lagoudakis2008,Krizhanovskii2010,Nardin2011,Boulier2015}. Moreover, the SO coupling induced by the TE-TM splitting \cite{Kavokin2005,Leyder2007,Hivet2012} allowed the observation of spin vortices and antivortices \cite{Manni2013,Sala2015,Dufferwiel2015}, which can bee seen as the eigenmodes of the four-dimensional first-excited manifold of a circular harmonic potential (two dimensions for the OAM $l=\pm 1$ and two for the polarisation degree of freedom). 

{\it Spin-Orbit Hyperspheres -} The pseudospin of photons can be represented by the Poincar\'{e} sphere, where each state can be seen as a coherent superposition of right ($\sigma^{+}$) and left ($\sigma^{-}$) circularly polarized light: $|\psi\rangle=\psi_{+}|\sigma^{+}\rangle+\psi_{-}|\sigma^{-}\rangle$, where $\psi_{+}$ and $\psi_{-}$ are complex numbers and $|\psi_{+}|^{2}+|\psi_{-}|^{2}=1$. The states $|\sigma^{+}\rangle$ and $|\sigma^{-}\rangle$ appear as the two poles of the Poincar\'{e} sphere having radius equal to 1. The positions of all other states on the sphere are determined by the differences of amplitude and phase between $\psi_{+}$ and $\psi_{-}$. Here we consider the wider Hilbert space made of all possible coherent superpositions of circularly polarized photons carrying OAM $l=\pm1$. A basis for this four dimensional space is:
\begin{equation}
\boldsymbol{\sigma}=(|\circlearrowleft,\sigma^{+}\rangle,|\circlearrowright,\sigma^{+}\rangle,|\circlearrowleft,\sigma^{-}\rangle,|\circlearrowright,\sigma^{-}\rangle),            
\label{basis1}
\end{equation}   
where $\circlearrowright$ and $\circlearrowleft$ represent $l=1$ and $l=-1$ OAM states respectively.  Each element in this basis is expressed in the form of $\sigma^{\pm}$-based Jones vectors:
\begin{equation}
\begin{aligned}
|\circlearrowleft,\sigma^{+}\rangle=C(r)\left( \begin{matrix}e^{-i\theta} \\ 0 \end{matrix} \right),\,\,\,\,\,\,\,\,\,
|\circlearrowright,\sigma^{+}\rangle=C(r)\left( \begin{matrix}e^{i\theta} \\ 0 \end{matrix} \right),\\
|\circlearrowleft,\sigma^{-}\rangle=C(r)\left( \begin{matrix} 0\\ e^{-i\theta} \end{matrix} \right),\,\,\,\,\,\,\,\,\,
|\circlearrowright,\sigma^{-}\rangle=C(r)\left( \begin{matrix} 0\\ e^{i\theta} \end{matrix}  \right),
\end{aligned} 
\label{basis2}
\end{equation}  
where the first (second) component of the column vector corresponds to the $\sigma^{+}$ ($\sigma^{-}$) polarisation, $\theta$ is the azimuthal angle in real space, and $C(r)$ is the radial intensity profile. These states can be viewed as poles of the hypersphere representing all the states $|\psi\rangle=\boldsymbol{\psi\cdot\sigma}$, with $\boldsymbol{\psi}=(\psi_{1},\psi_{2},\psi_{3},\psi_{4})$. Note that the hypersphere is identified by 6 parameters since of the 8 parameters corresponding to the 4 complex numbers $\psi_{1-4}$ only 6 are independent due to an arbitrary choice of a phase factor and to the condition $|\boldsymbol{\psi}|^{2}=1$.

States on the hypersphere involve OAM and pseudospin, and are thereby named: spin-orbit vectors (SOV). For any two orthogonal SOVs it is possible to generate spin-orbit Poincar\'{e} sphere (SOPS) using the same rules used to build the pseudospin Poincar\'{e} sphere. For example, the ``purely orbital" Poincar\'{e} sphere of Ref. \cite{Padgett1999} is generated by choosing $\boldsymbol{\psi}=(1,0,0,0)$ and $(0,1,0,0)$ as north and south poles, while the SOPSs studied in Refs. \cite{Milione2011,Milione2012} as an example of ``higher order" Poincar\'{e} spheres are generated by using the states: $\boldsymbol{\psi}=(1,0,0,0)$ and $(0,0,0,1)$ or $\boldsymbol{\psi}=(0,1,0,0)$ and $(0,0,1,0)$.

{\it The case of polaritons -} In Bragg-cavity polariton systems it is well known that Bragg reflectors introduce a splitting of the TE-TM modes that can be interpreted as an ``effective magnetic field" \cite{Kavokin2005,Flayac2013}. In the presence of this term the Hamiltonian of the system in the basis of the $\sigma^+$ and $\sigma^-$ lower-polariton modes is:
\begin{equation}
H=
\left(\begin{matrix}
\hbar\omega^{\sigma^+}_{LP}-\frac{\hbar^2\nabla^2}{2m_{LP}} + V & \beta\left(\frac{\partial}{\partial x} - i\frac{\partial}{\partial y}\right)^2
\\
\beta\left(\frac{\partial}{\partial x} + i\frac{\partial}{\partial y}\right)^2 & \hbar\omega^{\sigma^-}_{LP}-\frac{\hbar^2\nabla^2}{2m_{LP}} + V
\end{matrix}
\right),
\label{hamiltonian}
\end{equation}
where $\omega_{LP}^{\sigma^+/\sigma^-}$ are the polariton frequencies at normal incidence, $m_{LP}$ is the lower-polariton mass, and the terms depending on $\beta=\hbar^2(1/m_{t}-1/m_{l})/4$ describe the TE-TM splitting, where $m_{t/l}$ are the lower-polariton masses in the TE/TM polarisations. The harmonic confinement $V=\frac{1}{2}m_{LP}\omega^2_{HO}(x^2+y^2)$ can be experimentally realised using an open-cavity setup \cite{Dufferwiel2015}. The photon decay rate is simulated with non-Hermitian terms: $-i\gamma_{LP}/2$. Note that polariton-polariton interactions have been neglected, assuming that the polariton system is driven resonantly with weak laser pulses.

If $\beta=0$ this Hamiltonian reduces to the 2D harmonic potential for the two polarisations, and the Laguerre-Gauss modes with $l=\pm 1$ in the two polarisations are a basis for its first excited manifold, which is given by Eq.~(\ref{basis2}) with $C(r)=r\,e^{-r^2/2a^2}/\sqrt{\pi a^2}$ and $a=\sqrt{\hbar/m_{LP}\omega_{HO}}$. For small TE-TM splittings a perturbative approach may be used to demonstrate that the new system eigenmodes, for $\omega_{LP}^{\sigma^+}=\omega_{LP}^{\sigma^-}$, are: the two energy-split radial ($\boldsymbol{\psi_{RA}}=(1,0,0,1)/\sqrt{2}$) and azimuthal ($\boldsymbol{\psi_{AZ}}=(1,0,0,-1)/\sqrt{2}$) spin-vortices; and the two degenerate hyperbolic spin antivortices $\boldsymbol{\psi_{HY1}}=(0,1,1,0)/\sqrt{2}$ and $\boldsymbol{\psi_{HY2}}=(0,1,-1,0)/\sqrt{2}$  (see Fig.~\ref{fig1}(a)) \cite{Dufferwiel2015}. The two spin-vortices energy  are $E=E_1\pm 2\beta/a^2$ while the spin antivortices energy is $E=E_1$, where $E_1$ is the energy of the unperturbed modes (equal for the two $\sigma^{\pm}$ components). The states with higher OAM exhibit much higher energies and do not influence the dynamics of the OAM=1 states and will be neglected in the following.
\begin{figure}
\center
\includegraphics[scale=0.35]{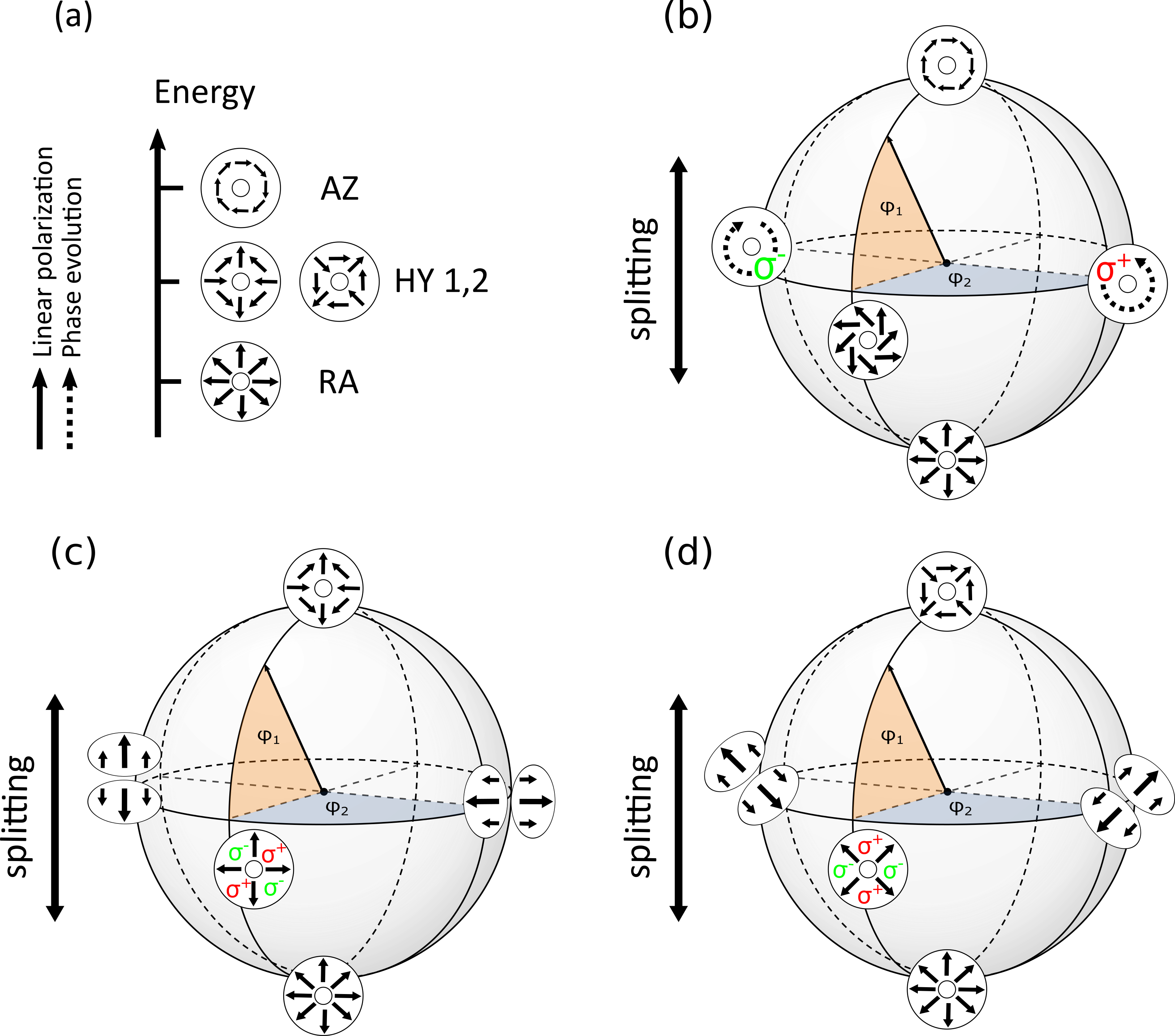}
\center
\caption{\label{fig1} Schematic of the polariton states. (a) eigenenergies and eigenmodes in the presence of 2D harmonic confinement and TE-TM splitting. (b-d): SO-Poincar\'e spheres resulting from the states: $\boldsymbol{\psi_{RA}}$ and  $\boldsymbol{\psi_{AZ}}$ (b), $\boldsymbol{\psi_{RA}}$ and $\boldsymbol{\psi_{HY1}}$ (c), and $\boldsymbol{\psi_{RA}}$ and $\boldsymbol{\psi_{HY2}}$ (d) (detailed mathematical description in Supplementary). The angular coordinates are: $\varphi_1$ and $\varphi_2$.}
\end{figure}

{\it State manipulation -} To manipulate the SOV state it is possible to take advantage of the energy structure of the perturbed eigenmodes and of the AC Stark effect. The energy structure of the new eigenmodes suggests a decomposition of the SO hypersphere into three SOPSs with an energy splitting between the north and south poles (Fig.~\ref{fig1}). This splitting acts as an {\it effective} steady state magnetic field \cite{Kavokin2005} inducing the precession of a SOV state around the vertical axis of a sphere. We note here that the 6D hypersphere can be decomposed in several 2D spheres, together with those in Fig.~\ref{fig1}, but these additional spheres do not influence the state manipulation. Moreover, due to the AC Stark effect, a laser pulse far-red detuned from an exciton line induces a transient blue-shift of the exciton resonance with the same polarisation. Recently, it has been shown that if the exciton is strongly coupled to a photon mode, the blue shift is transferred to a blue-shift of the polariton lines \cite{PhysRevLett.109.033605}. Therefore, a Stark pulse $\sigma^+$ polarised will induce a splitting between the $\sigma^+$ and $\sigma^-$ polaritons that will act as an {\it effective} pulsed magnetic field. Therefore, in the case of the SOPS in Fig.~\ref{fig1}(b) the splitting between the $\sigma^+$ and $\sigma^-$ components will induce a rotation of the state around the axis connecting the two $\sigma^{\pm}$ phase vortices at the equator lasting for the length of the Stark pulse. Similarly, linearly polarised pulses will induce rotations around the horizontal axes at the equators of the SOPSs of Fig.~\ref{fig1}(c) and (d). As the Stark pulse is red-detuned with respect to the exciton line it does not inject any polariton in the cavity.

To evaluate the dynamic of the relevant states, for example under the effect of $\sigma^{\pm}$ Stark pulses, it is sufficient to solve the system of linear equations defined by the following matrix:
\begin{equation}
\left(
\begin{matrix}
\hbar\delta\omega^{\sigma^+}_{LP}(t) & 0 & 0 & -2\beta/a^2 \\
0 & \hbar\delta\omega^{\sigma^+}_{LP}(t) & 0 & 0 \\
0 & 0 & \hbar\delta\omega^{\sigma^-}_{LP}(t) & 0 \\
-2\beta/a^2 & 0 & 0 & \hbar\delta\omega^{\sigma^-}_{LP}(t) \\
\end{matrix}
\right),
\end{equation}
\noindent
where $\hbar\delta\omega^{\sigma^{\pm}}_{LP}(t)=\hbar\delta\omega^{\sigma^{\pm}}_{LP}exp[-(t-t^{st})^2/2\sigma_{st}^2]$ describes the time dependent Stark shift of the $\sigma^{\pm}$ polaritons, centred at $t^{st}$ and with width $\sigma_{st}$ (see Supplementary Material for $H$, $V$, $D^+$, and $D^-$ polarised pulses). The eigenenergies for the system (i.e. $E_1+\hbar\delta\omega^{\sigma^+}_{LP}$, $E_1+\hbar\delta\omega^{\sigma^-}_{LP}$, and $\frac{\hbar}{2}\left(\delta\omega^{\sigma^+}_{LP}+\delta\omega^{\sigma^+}_{LP}\pm\sqrt{(\delta\omega^{\sigma^+}_{LP}-\delta\omega^{\sigma^-}_{LP})^2+16\beta^2/a^2\hbar^2}\right)$, where the $t$ dependence has been omitted) show that the splitting between the $\sigma^{\pm}$ polaritons is mapped to a transient shifting of the system eigenmodes.

\begin{figure}
\includegraphics[scale=0.325]{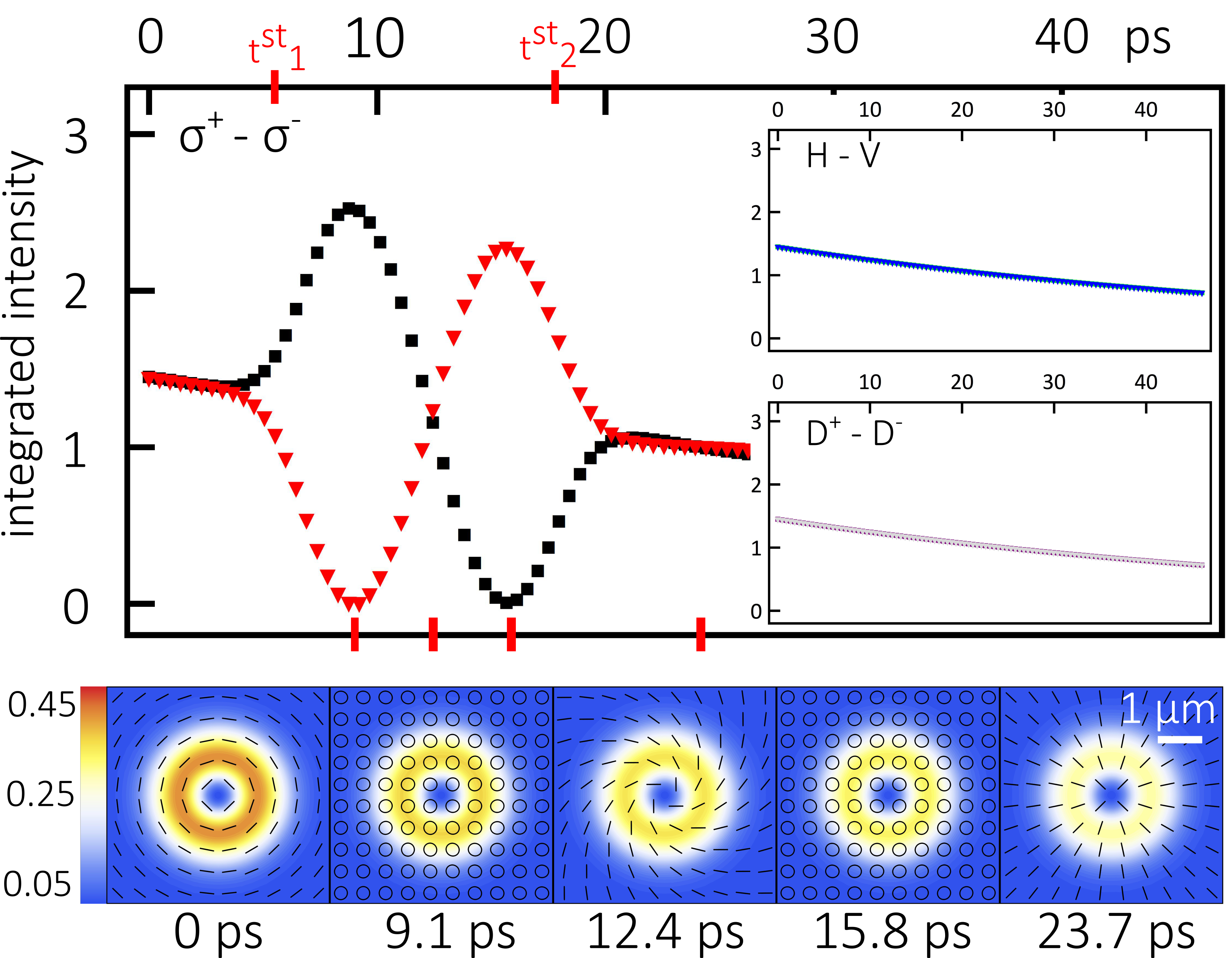}
\caption{\label{fig2} Manipulation of a SOV state on the SOPS in Fig.~\ref{fig1}(b), starting from $\boldsymbol{\psi_{AZ}}$ to arrive to $\boldsymbol{\psi_{RA}}$ (animation in Supplementary material). Top panel: polariton population as a function of time: $\sigma^+$ (black squares), $\sigma^-$ (red triangles) in the main panel, $H$ (green squares), $V$ (blue triangles) and $D_+$ (purple squares), $D_-$ (grey triangles) in the insets. Bottom panel: polariton distribution and polarisation at different times, the colour code indicates the local polariton density. Two $\sigma^+$ polarised Stark pulses are simulated by a shift of the $\sigma^+$ polariton with $\sigma_{st}=1$ ps and $\hbar\delta\omega^{\sigma^+}=0.4$ $meV$, and centred at $t^{st}_1=6.58$ and $t^{st}_2=18.15$ ps. The parameters here and in the rest of the paper are:  $m_{LP}=2.4\times10^{-5}m_e$ ($m_e$ being the electron mass), $\omega_{HO}=4.0$ $ps^{-1}$, $\beta=0.04$ $meV\mu m^2$, $\gamma_{LP}=0.02$ meV \cite{PhysRevX.3.041015,Dufferwiel2014}.}
\end{figure}
To demonstrate the complete control over a SOV state it is sufficient to demonstrate its manipulation on the SOPSs of Fig.~\ref{fig1} and show that is possible to move it from a given eigenstate of the basis to all the other three. As a first example, Fig.~\ref{fig2} shows the polarisation-resolved density and spatial polariton distribution and polarisation during a manipulation from the $\boldsymbol{\psi_{AZ}}$ state (north pole) to the $\boldsymbol{\psi_{RA}}$ state (south pole) of the SOPS in Fig.~\ref{fig1}(b). Note that here we do not address the case of single-polariton/photon operations.

This can be achieved using two $\sigma^+$-polarised Stark pulses. A first pulse, arriving at $t=t^{st}_1=6.58$ ps, flips the SOV from the north pole to the equator where, due to the energy spitting between the two poles, it will precess passing from a $\sigma^+$-polarised to a $\sigma^-$-polarised vortex. Then a second pulse, arriving at $t=t^{st}_2=18.15$ ps, flips the SOV from the equator to the $\boldsymbol{\psi_{RA}}$ state. In Fig.~\ref{fig2} the system is initialised ($t=0$ ps) in the $\boldsymbol{\psi_{AZ}}$ state, as can be seen by observing that the $\sigma^+$ and $\sigma^-$ components have the same intensity and analysing the polariton distribution and polarisation in the lower panel. Between the two Stark pulses ($t^{st}_1<t<t^{st}_2$) the SOV state precesses passing from a $\sigma^+$-polarised to a spiral and then to a $\sigma^-$-polarised state. This can be seen from the oscillations of the $\sigma^+$ and $\sigma^-$ populations, and from the polarisation structure in the lower panel at $t=9.1,12.4,15.8$ ps. Finally, after the second $\sigma^+$-polarised Stark pulse, arriving when the SOV is half-way between the $\sigma^-$ and the $\sigma^+$ vortex (i.e. at $\varphi_{2}=3\pi/2$ as defined in Fig.~\ref{fig1}), the SOV is at the south pole. This can be seen by the $\sigma^{\pm}$ components being again balanced, and by the spatial polariton distribution and polarisation at $t=23.7$ ps.

It is worth mentioning here that the same manipulation of the SOV can be achieved in other ways. For example, two $\sigma^+$ pulses with different intensities can be used to flip the SOV state first to a nonzero latitude (not the equator) and then to the south pole. Alternatively, two pulses with $\sigma^-$ polarisation or two pulses with opposite circular polarisations could have been used. This can be understood by observing that $\sigma^+$ and $\sigma^-$ polarised pulses induce rotations in opposite directions: a $\sigma^-$ polarised pulse arriving when the SOV state is at $\varphi_{2}=3\pi/2$ flips it to the north pole, not to the south pole as done by a $\sigma^+$ pulse in Fig.~\ref{fig2}. Instead the same $\sigma^-$ polarised pulse arriving when the SOV state is at $\varphi_{2}=\pi/2$ flips it to the south pole.

It is worth noticing here that this manipulation has been achieved simply using Stark pulses with $\sigma^{\pm}$ polarisation without any requirement of spatial structure or OAM. This is because each SOV state on the sphere is a linear combination of $\sigma^+$ and $\sigma^-$ components with a relative weight that varies along the horizontal axis of the sphere from $\varphi_2=0$ (pure $\sigma^+$) to $\varphi_2=\pi$ (pure $\sigma^-$). Therefore, Stark pulses simply $\sigma^{+}$ or $\sigma^{-}$ polarised are enough to split the energy of the $\varphi_1=\pi/2,\varphi_2=0$ and $\varphi_1=\pi/2,\varphi_2=\pi$ points and to induce a rotation of a SOV state around this axis. Instead, linearly polarised Stark pulses will have no effect on the SOVs on this sphere since they all exhibit the same fraction of linearly polarised components at any $\varphi_1$ and $\varphi_2$. Similarly on the SOPS with $\boldsymbol{\psi_{RA}}$ and $\boldsymbol{\psi_{HY1}}$ as poles [Fig.~\ref{fig1}(c)] all the states are linear combination of H-V polarisations and $H$ and $V$ polarised pulses can be used to manipulate the states on this sphere. In the same way the manipulation of the SOV state on the SOPS with $\boldsymbol{\psi_{RA}}$ and $\boldsymbol{\psi_{HY2}}$ as south and north poles [Fig.~\ref{fig1}(d)] can be achieved by means of $D^+$ and $D^-$ polarised pulses. Therefore, SOV states can be manipulated independently on the three spheres using pulses with different polarisations. Note that the efficiency of this selective exciton shift is strongly dependent on the exciton confinement (3D, 2D or 1D) and on the material \cite{PhysRevB.65.155206,0953-8984-2-27-005}.

\begin{figure}
\includegraphics[scale=0.325]{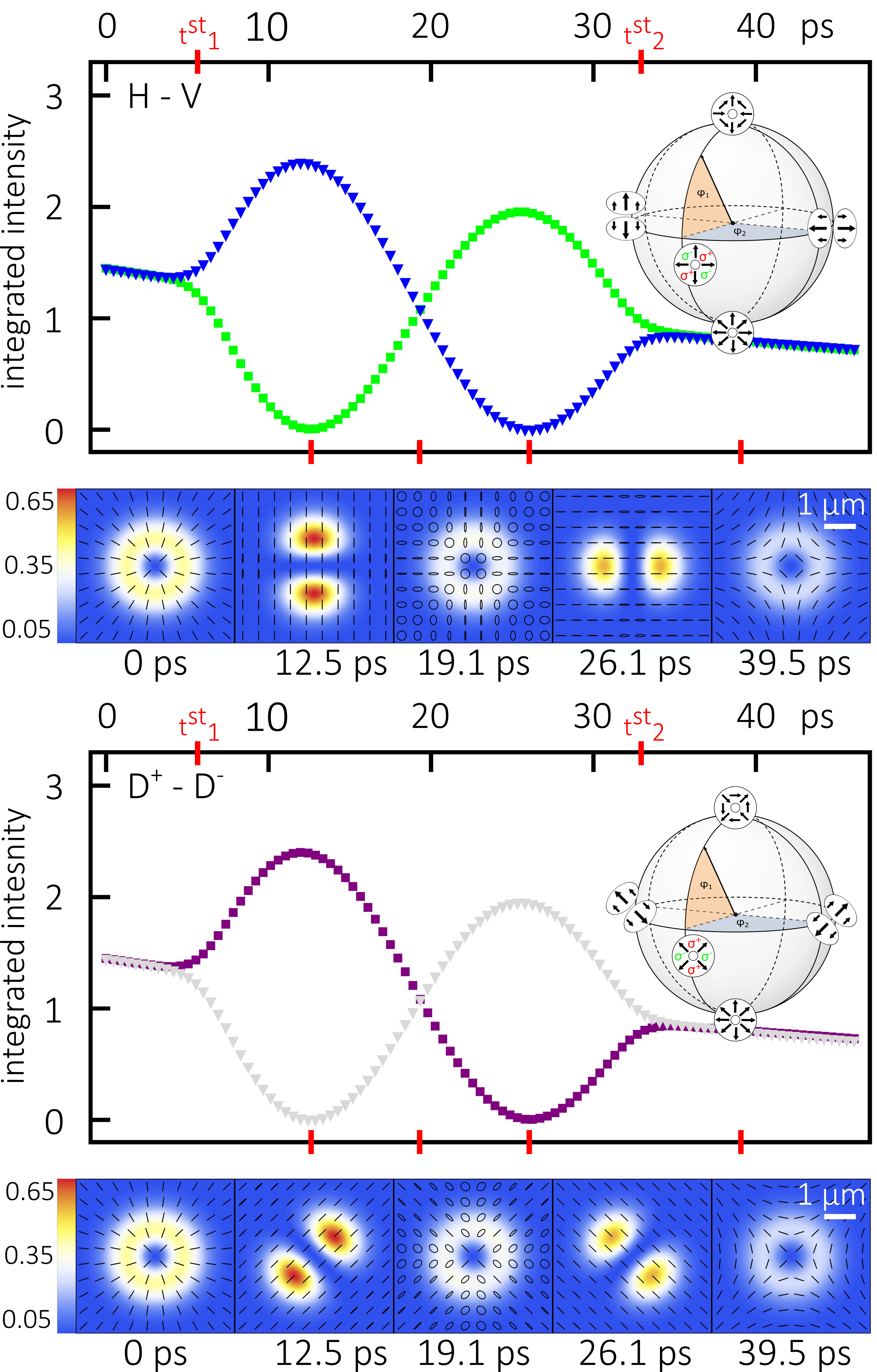}
\caption{\label{fig3} Polariton population as a function of time and polariton distribution and polarisation during a SOV manipulation (animations in Supplementary material). Top panels: manipulation on the SOPS of Fig.~\ref{fig1}(c) with the $H$ and $V$ polarisations in green squares and blue triangles. Bottom panels: manipulation on the SOPS of Fig.~\ref{fig1}(d) with the $D^+$ and $D^-$ polarisations in purple squares and grey triangles. The Stark pulses are simulated by two consecutive shifts of the $H$ (respectively $D^+$) polarised polariton with $\sigma_{st}=1$ ps, $\delta\omega^{H}=\delta\omega^{D^+}=0.305$ $meV$, $t^{st}_1=6.58$ and $t^{st}_2=32.1$ ps.}
\end{figure}

The top panels of Fig.~\ref{fig3} address the case of SOPS in Fig.~\ref{fig1}(c) formed by the $\boldsymbol{\psi_{RA}}$ and the $\boldsymbol{\psi_{HY1}}$ states. An $H$ polarised pulse flips the state from the south pole to the equator where, due to the energy separation between the two poles (Fig.~\ref{fig1}(a)), the SVO starts to precess. Comparing Fig.~\ref{fig2} with Fig.~\ref{fig3}, it is clear that the precession is much slower in this second case, since the energy separation between the poles is smaller. More precisely, the separation between the $\boldsymbol{\psi_{AZ}}$ and the $\boldsymbol{\psi_{RA}}$ states [Fig.~\ref{fig1}(b) and Fig.~\ref{fig2}] is $\Delta E=0.306$ meV, which corresponds to a period of $13.4$ ps. Instead the separation between $\boldsymbol{\psi_{RA}}$ and $\boldsymbol{\psi_{HY1}}$ [Fig.~\ref{fig1}(c) and Fig.~\ref{fig3}] is $\Delta E=0.15$ meV, which corresponds to a period of $26.8$ ps. As before, when the SOV rotating at the equator reaches the mid point ($\varphi_2=3\pi/2$) between the $H$- and $V$-polarised states, a second shift of the $H$-polarised exciton rotates the SOV state to the north pole $\boldsymbol{\psi_{HY1}}$.

The case of the SOPS with $\boldsymbol{\psi_{RA}}$ and $\boldsymbol{\psi_{HY2}}$ as poles [Fig.~\ref{fig1}(d)] is very similar but diagonally polarised pulses are needed to manipulate the system. In analogy with what is seen in the insets of Fig.~\ref{fig2}, the polarisation components which are not affected, namely circularly and diagonally (resp. horizontally-vertically) in the SOPSs for Fig.~\ref{fig1}(c) [resp. Fig.~\ref{fig1}(d)], trivially decrease monotonically in time (not shown here). Note that since the three considered SOPSs have $\boldsymbol{\psi_{RA}}$ as south pole, it is possible to move the SOV state from one sphere to another by applying Stark pulses with different polarisations.

Finally, we tested these manipulations using different parameters (variations up to 10$\%$ in the values of $\sigma_{st}$, and $\delta\omega$ were considered). While the range of parameters for which the manipulations can be achieved is quite broad, the relation between them is particularly critical, especially for the amplitudes, durations and times of the energy shifts. Increased values of TE-TM splitting $\beta$ (see Supplementary material for the case $\beta=0.06$ $meV\mu m^2$) and longer polariton lifetimes allow for a higher number of manipulations. Note that, in order to reach the level of quantum information processing dissipation needs to be low enough to have a well defined number of particles present inside the cavity for the entire duration of the manipulation. For our work we adopted a semiclassical description of the polariton system, which is justified by the fact that the coherence time of resonantly pumped polaritons systems is longer than the photon lifetime inside the cavity.

To conclude, we proposed a model based on a hypersphere to study the evolution of spin-orbit vector states. We have demonstrated that thanks to the hybrid nature of dressed half-light half-matter states spin vortices can be efficiently manipulated by means of red-detuned Stark pulses with different polarisations. This model valid in the presence of strong light-matter coupling, circular confinement, and TE-TM splitting allows the manipulation of multiple individual logical bits within one physical system and to generalise to SO-coupled states with OAM larger than 1. Moreover in the case of semiconductor open-cavity systems this model is already within experimental feasibility \cite{PhysRevLett.109.033605,Dufferwiel2015}. In order to implement quantum information protocols based on single-polaritons/photons operations, microcavities can be coupled to external single photon sources \cite{PhysRevLett.115.196402} and the pseudospin manipulation performed faster than the polariton lifetime. Our approach, in the case of single-photon systems, can lead to the implementation of a new type of quantum electrodynamics based on spin-orbit coupled states. In addition, it can also be an efficient method to manipulate the OAM and spin of a polariton condensate

\begin{acknowledgements}
We acknowledge support by EPSRC grant EP/J007544, ERC Advanced Grant No. EXCIPOL 320570 and the Leverhulme Trust Grant No. PRG-2013-339.
\end{acknowledgements}

\bibliography{manuscript}

\newpage
\begin{widetext}
\section{Supplementary: Effect of Stark Pulses With Different Polarisations}

We present here the theoretical approach used to simulate the dynamic of the system under the effect of red-detuned Stark pulses. With respect to the main text, where we present only the case of $\sigma^+$ and $\sigma^-$ polarised pulses, here we also give the results for pulses $H$, $V$, $D^+$ and $D^-$ polarised.

To study the dynamic of the system it is possible to take advantage of the fact that the effect of a far red-detuned Stark pulse, with a given polarisation, is an almost instantaneous blue shift of the polariton energy with the same polarisation \cite{PhysRevLett.109.033605,PhysRevLett.112.053601}. Therefore, it is possible to account for the effect of the pulse simply by mapping its time-profile and intensity into time-dependent ``effective'' eigenenergies for the system. More technically, it is possible to use a time-dependent perturbative approach to derive the eigenmodes of the system for different small variations of the bare polariton energy (i.e. the polariton energy without harmonic confinement), and then study the evolution of the system with eigenmodes that vary in time following the time-profile and intensity of a Stark pulse.

The Hermitian part of the system's Hamiltonian can be written, in the base of the $\sigma^+$ and $\sigma^-$ lower-polariton modes, as:
\begin{equation}
H=
\left(\begin{matrix}
\omega^{\sigma^+}_{LP}-\frac{\hbar^2\nabla^2}{2m_{LP}} + V & \beta\left(\frac{\partial}{\partial x} - i\frac{\partial}{\partial y}\right)^2
\\
\beta\left(\frac{\partial}{\partial x} + i\frac{\partial}{\partial y}\right)^2 & \omega^{\sigma^-}_{LP}-\frac{\hbar^2\nabla^2}{2m_{LP}} + V 
\end{matrix}
\right),
\label{eq1}
\end{equation}
where $\omega_{LP}^{\sigma^+}$ ($\omega_{LP}^{\sigma^-}$) is the $\sigma^+$ ($\sigma^-$) polariton frequency at normal incidence, $m_{LP}$ is the effective mass of the lower-polariton, $V=\frac{1}{2}m_{LP}\omega^2_{HO}(x^2+y^2)$ is the harmonic confinement, and the terms depending on $\beta=\hbar^2(1/m_{t}-1/m_{l})/4$ describe the TE-TM splitting, where $m_{t/l}$ are the lower-polariton masses in the TE/TM polarizations (as in the main text the system is considered in the linear regime and polariton-polariton interactions are neglected). In the case of zero TE-TM splitting this Hamiltonian reduces to the quantum harmonic oscillators, and the eigenmodes of its first excited manifold are the Laguerre-Gauss modes with $l=\pm 1$ in the two polarisations, which are equal to the Jones vectors defined in the main text:

\begin{equation}
\begin{aligned}
|1\rangle=|\circlearrowleft,\sigma^{+}\rangle=C(r)\left( \begin{matrix}e^{-i\theta} \\ 0 \end{matrix} \right),\,\,\,\,\, 
|2\rangle=|\circlearrowright,\sigma^{+}\rangle=C(r)\left( \begin{matrix}e^{i\theta} \\ 0 \end{matrix} \right),\\
|3\rangle=|\circlearrowleft,\sigma^{-}\rangle=C(r)\left( \begin{matrix} 0\\ e^{-i\theta} \end{matrix} \right),\,\,\,\,\,
|4\rangle=|\circlearrowright,\sigma^{-}\rangle=C(r)\left( \begin{matrix} 0\\ e^{i\theta} \end{matrix}  \right),
\end{aligned}           
\end{equation}  

\noindent
with $C(r)=r\,e^{-r^2/2\sigma^2}/\sqrt{\pi\sigma^2}$ and $\sigma=\sqrt{\hbar/m_{LP}\omega_{HO}}$).

As previously shown, the effect of TE-TM splitting, in the case of strong harmonic confinement, can efficiently be treated perturbatively \cite{Dufferwiel2015}. For this reason, in order to simulate the dynamic of the system we use a time-dependent perturbative approach for both the TE-TM splitting and the shift of the polariton line induced by a Stark pulse. Using the base just defined, the perturbation matrix $M$ (with elements $M_{i,j}=\langle i| H |j\rangle$, with $i,j=1,...,4$) in the presence of a shift of the circularly polarised polariton lines is:

\begin{equation*}
\left(
\begin{matrix}
\delta\omega^{\sigma^+}_{LP}(t) & 0 & 0 & -2m_{LP}\omega_{HO}\beta \\
0 & \delta\omega^{\sigma^+}_{LP}(t) & 0 & 0 \\
0 & 0 & \delta\omega^{\sigma^-}_{LP}(t) & 0 \\
-2m_{LP}\omega_{HO}\beta & 0 & 0 & \delta\omega^{\sigma^-}_{LP}(t) \\
\end{matrix}
\right).
\end{equation*}

\noindent
where $\delta\omega^{\sigma^{\pm}}_{LP}(t)=\delta\omega^{\sigma^{\pm}}_{LP}\,exp[-(t-t_{st})^2/2\sigma_{st}^2]$ describe the modifications to the $\sigma^{\pm}$ polarised polaritons induced by the Stark pulses. This matrix defines a set of linear differential equations describing the time evolution of the system in the presence of $\sigma^{\pm}$-polarised Stark pulses, which can be evaluated numerically for each time. The corresponding matrices for Stark pulses polarised $H$ and $V$, and $D^+$ and $D^-$ are:

\begin{equation*}
\left(
\begin{matrix}
\frac{\delta\omega^{H}_{LP}(t)+\delta\omega^{V}_{LP}(t)}{2} & \frac{\delta\omega^{H}_{LP}(t)-\delta\omega^{V}_{LP}(t)}{2} & 0 & -2m_{LP}\omega_{HO}\beta \\
\frac{\delta\omega^{H}_{LP}(t)-\delta\omega^{V}_{LP}(t)}{2} & \frac{\delta\omega^{H}_{LP}(t)+\delta\omega^{V}_{LP}(t)}{2} & 0 & 0 \\
0 & 0 & \frac{\delta\omega^{H}_{LP}(t)+\delta\omega^{V}_{LP}(t)}{2} & \frac{\delta\omega^{H}_{LP}(t)-\delta\omega^{V}_{LP}(t)}{2} \\
-2m_{LP}\omega_{HO}\beta & 0 & \frac{\delta\omega^{H}_{LP}(t)-\delta\omega^{V}_{LP}(t)}{2} & \frac{\delta\omega^{H}_{LP}(t)+\delta\omega^{V}_{LP}(t)}{2} \\
\end{matrix}
\right).
\end{equation*}

\noindent
and

\begin{equation*}
\left(
\begin{matrix}
\frac{\delta\omega^{D^+}_{LP}(t)+\delta\omega^{D^-}_{LP}(t)}{2} & i\frac{\delta\omega^{D^+}_{LP}(t)-\delta\omega^{D^-}_{LP}(t)}{2} & 0 & -2m_{LP}\omega_{HO}\beta \\
-i\frac{\delta\omega^{D^+}_{LP}(t)-\delta\omega^{D^-}_{LP}(t)}{2} & \frac{\delta\omega^{D^+}_{LP}(t)+\delta\omega^{D^-}_{LP}(t)}{2} & 0 & 0 \\
0 & 0 & \frac{\delta\omega^{D^+}_{LP}(t)+\delta\omega^{D^-}_{LP}(t)}{2} & i\frac{\delta\omega^{D^+}_{LP}(t)-\delta\omega^{D^-}_{LP}(t)}{2} \\
-2m_{LP}\omega_{HO}\beta & 0 & -i\frac{\delta\omega^{D^+}_{LP}(t)-\delta\omega^{D^-}_{LP}(t)}{2} & \frac{\delta\omega^{D^+}_{LP}(t)+\delta\omega^{D^-}_{LP}(t)}{2} \\
\end{matrix}
\right).
\end{equation*}

\section{Supplementary: Manipulation with higher TE-TM splitting ($\beta=0.06$ $meV\mu m^2$)}
We present here the results for the same manipulations performed in the main text but for a different, increased, value of the TE-TM splitting. While in the main text the value $\beta=0.04$ $meV\mu m^2$ was used, here we use the value $\beta=0.06$ $meV\mu m^2$. The increased value of the TE-TM splitting has two main consequences on the manipulation. First, a higher Stark shift is needed to perform the manipulation since the energy levels are more far apart. Second, the precession around the equator of a sphere is faster and therefore each manipulation can be performed on shorter time scales. In the case considered here the manipulation has been achieved in $10$ $ps$ for Fig.\ref{fig1} (instead of the $15$ $ps$ needed in the main text case) and in $20$ $ps$ for Fig.\ref{fig2} (instead of the $30$ $ps$ needed in the main text case).

\begin{figure}[h]
\includegraphics[scale=0.3]{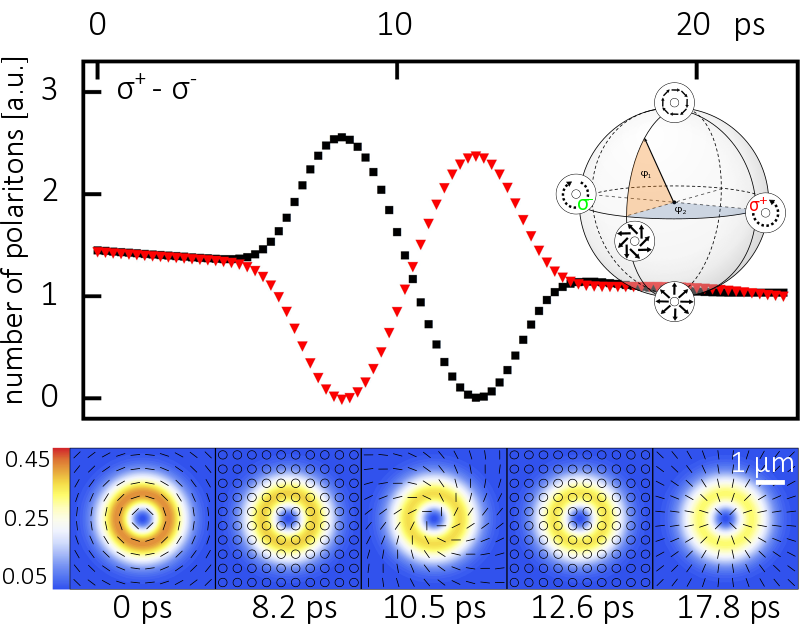}
\caption{\label{fig1} Manipulation of a SOV state. The system is initialised in the $\boldsymbol{\psi_{AZ}}$ and moved to the $\boldsymbol{\psi_{RA}}$ state. Top panel: polariton population as a function of time in the $\sigma^+$ (black squares) and $\sigma^-$ (red triangles) polarisations. Bottom panel: polariton distribution and polarisation at different times: t=0, 8.2, 10.5, 12.6, 17.8 ps (the colour code indicates the local number of polaritons). The state modulation is performed by means of two $\sigma^+$ polarised Stark pulses, which are simulated by a shift of the $\sigma^+$-polarised polariton with $\sigma_{st}=0.7$ ps and $\hbar\delta\omega^{\sigma^+}=0.6$ $meV$, and centred at $t_{st}^1=6.58$ and $t_{st}^2=14.33$ ps. The parameters used here are:  $m_{LP}=2.4\times10^{-5}m_e$ ($m_e$ being the electron mass), $\omega_{HO}=4.0$ $ps^{-1}$, $\beta=0.06$ $meV\mu m^2$, $\gamma_{LP}=0.02$ meV.}
\end{figure}

\begin{figure}
\includegraphics[scale=0.3]{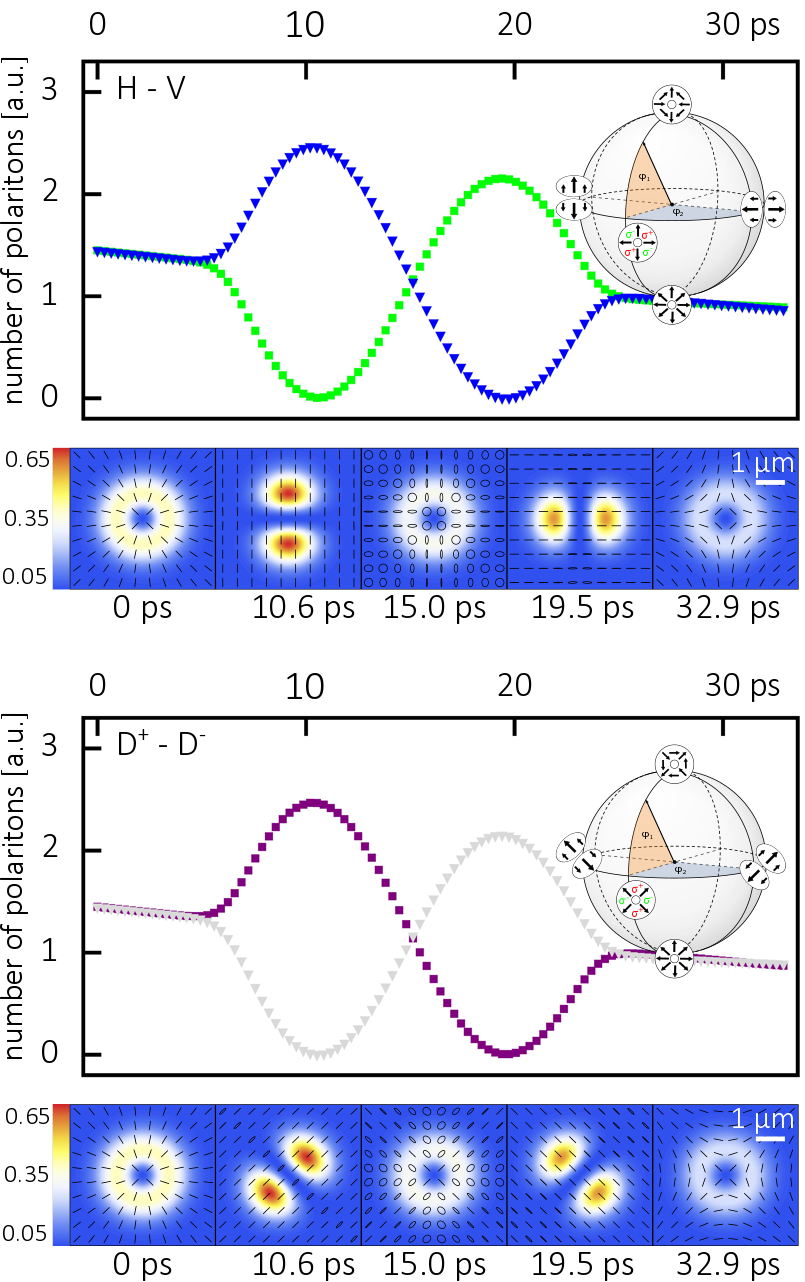}
\caption{\label{fig2} Manipulation of a SOV state. Top panels: the system is initialised in the $\boldsymbol{\psi_{RA}}$ and moved to the $HY1$ state. The two top panels show: the polariton population in the $H$ and $V$ polarisations (green squares and blue triangles) and the population and polarisation spatial distributions as a function of time. Bottom panels: the system is initialised in the $\boldsymbol{\psi_{RA}}$ and moved to the $HY2$. The two bottom panels show: the polariton population in the $D^+$ and $D^-$ polarisations (purple squares and grey triangles) and the population and polarisation spatial distributions as a function of time. In both cases the Stark pulses are simulated by two consecutive shifts of the $H$ (respectively $D^+$) polarised polariton lines with $\sigma_{st}=0.7$ ps, $\delta\omega^{H}=\delta\omega^{D^+}=0.46$ $meV$, $t_{st}^1=6.58$ and $t_{st}^2=23.65$ ps.}
\end{figure}

\newpage
\end{widetext}

\end{document}